\documentclass[
$$    ,final            
  ]
  {aipproc}

\layoutstyle{6x9}

\begin{document}
\title{ The ${\bar N}N $ quasi-bound states:  J$/\psi$ and atomic evidence\footnote{Invited talk (S. Wycech) to the International Conference on LOW ENERGY ANTIPROTON PHYSICS (LEAP05), May 16-22, 2005, Bonn-J\"{u}lich, Germany to appear in AIP series of conference proceedings.} }

\author{S{\l}awomir Wycech}{ 
  address={So{\l}tan Institute for Nuclear Studies,Warsaw, Poland}
  }

\author{ Benoit Loiseau }{
 address={Laboratoire de Physique Nucl\'eaire et de Hautes \'Energies,
Groupe Th\'eorie, Univ. P. et M. Curie, Paris, France}
 }


\begin{abstract}

The measurements  of J/$\psi$ decays into $
\gamma p {\bar p}$ show a strong enhancement at $ p {\bar p}$
threshold  not seen in the decays into  $ \pi^0 p {\bar p}$. 
What is the nature of this enhancement? 
A natural interpretation can be performed in terms of a classical
model of $ N \bar{N} $ interactions based on $G$-parity  transformation. 
The observed $ p {\bar p}$ structure is the consequence of the  strong 
attraction in the  $^{1}S_{0} $ state related predominantly to  
$\pi$-meson exchanges. Similar attractions generate near threshold:
 a virtual (or quasi-bound) 
state in  $^{11}S_{0}$-, a quasi-bound state in $^{33}P_{1} $-
and a resonance in $^{13}P_{0} $-waves. 
These    $P$-wave  structures find 
support in the $ {\bar p}$-atomic data. 

\end{abstract}

\maketitle


\section{ Introduction}

An  old question  in the antiprotonic physics is the  existence 
or non-existence  
of exotic  $N - \bar{N}$  systems: 
quasi-bound, virtual,  resonant, multiquark or baryonium states~\cite{kle02}.
Such states, if located close to the threshold,  would generate 
large  scattering lengths for  a given spin and isospin state. 
The scattering 
experiments offer  the easiest check   but a clear separation of   
quantum states  is not  easy.  Complementary measurements of  the 
X-ray transitions in  ${\bar p}$ atoms may select  
certain partial waves when the level  fine structure is resolved.  
This resolution was  achieved in the 1S states~\cite{aug99}   
and partly in the 2P states~\cite{augnp99} of hydrogen.  
Another method   to reach selected states are  formation experiments. 
Along this way, the  BES Collaboration~\cite{bai03} measured  the decays 
\begin{equation}
\label{1} 
J/\psi \rightarrow \gamma p\bar{p}
\end{equation} 
and found an  enhancement close to the $ p\bar{p} $ 
threshold.  
A  clear threshold suppression is seen in another 
decay channel $J/\psi \rightarrow  \pi^0  p\bar{p}$.
Conservation  laws   limit  quantum numbers of  the 
$ p\bar{p}$ states allowed in those decays. The final state $ p\bar{p}$ 
interactions reduce these further to one state per channel. 
While the $\bar{p}$-hydrogen  determines   scattering lengths (volumes), 
the  $J/\psi$ decays allow  to extend this knowledge  to energies 
above the threshold. To look  below the $ p\bar{p}$ threshold 
one needs heavier $ \bar{p}$ atoms.

Two studies are presented  in this contribution:  

$\bullet$ The  $J/\psi$ decay mode  (\ref{1}) is discussed and the 
threshold $p \bar{p}$ enhancement is attributed to 
a broad subthreshold state in the $ ^{11}S_0 $ wave. 
 
$\bullet$ 
The atomic level  shifts are related to 
the $\bar{p}$-nucleus  zero energy  scattering parameters $A_L$.  
In light   atoms the latter  are 
extracted from  the  $\bar{p}d$-, $\bar{p} ^3$He-, $\bar{p} ^4$He-data. 
Next, $A_L$ are  expressed  in terms of  
the $\bar{p} p$-, $\bar{p}n$-subthreshold lengths $a(E)$ and volumes 
$b(E)$.  Due to the differences in nuclear binding 
one can  obtain, in this way, the energy dependence 
of  Im $a(E)$, Im $b(E)$.This dependence  indicates  a $P$-wave quasi-bound state.


\section{The $\bar{p}p $ final state interactions in $ J/\psi$  decays}

\begin{figure}
\includegraphics[height=.38\textheight]{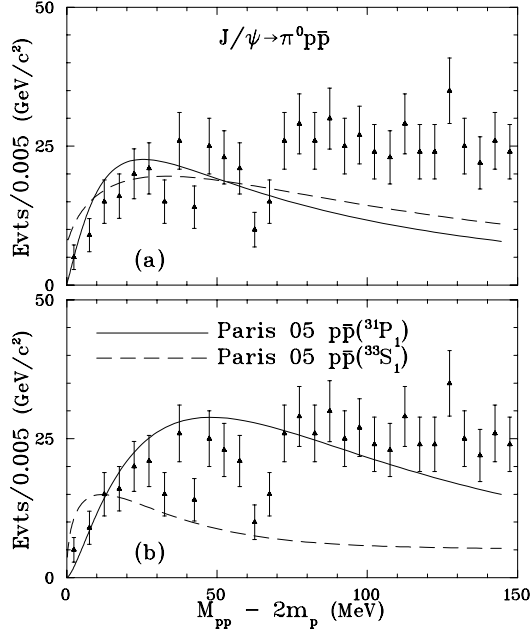}
\caption{ The $ \pi^0  p \bar{p} $ decay channel.
The experimental data has been extracted from Fig.~2(a) 
of Ref.~\protect\cite{bai03}. The solid and dashed lines represent the results obtained with the  
$^{31}P_1$- and $^{33}S_1$-waves of the recent version~\cite{par05} 
of the Paris potential, respectively. The previous version of the Paris
potential~\cite{par99} give similar results.  (a) The  final state factor 
$q\mid T_{ff}/q^{L} \mid^2$ (Watson approximation).
The constant $C_{if}$  is chosen to fit  the  
low-energy part of the  data. 
This approximation  fails  for $M_{p\bar{p}}-2m_p >40$~MeV ($ q>1$~fm$^{-1}$). 
(b) The  rate $q\mid T_{if}\mid^2$  of Eq.~(\ref{2}). 
The constant $A^0_{if}$
and the formation range parameter $r_o=0.55$~fm 
are  chosen to obtain a  good fit to the data. Here, the  $^{31}P_1$ wave
reproduces well the data.
A 18~MeV wide  state bound by  
18~MeV  is  generated 
with the Paris model~\cite{par05}   in the $^{31}P_1$ wave, but 
it has little effect on the results. }    
\end{figure}

\begin{figure}
\includegraphics[height=.38\textheight]{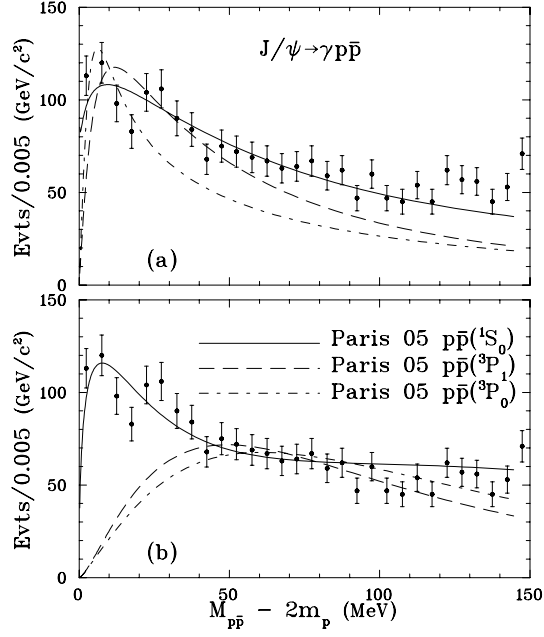}
\caption{The $ \gamma  p \bar{p} $ decays. 
Data as in Fig.~1. The solid, dashed and dot-dashed lines represent the 
results obtained with the  $^{1}S_0$-,
$^{3}P_1$- and $^{3}P_0$-waves of~\cite{par05}, respectively. 
(a)  The final state factor
$q\mid T_{ff}/q^L\mid^2$ (Watson approximation).  
At $q >2~$ fm$^{-1}$this approximation begins to fail.
(b) The rate  \protect\linebreak $q\mid T_{if}\mid^2$ of Eq.~(\ref{2}) with  $r_o=0.55$~fm. 
The $ ^{1}S_0 $-wave of~\protect \cite{par05} 
offers the best fit to the  data. It involves a  quasi-bound state in the $ ^{1}S_0 $-wave   
located very close to threshold, of  $ 53 $~MeV width  and 
$ 5 $~MeV binding. } 
\end{figure}

The $J^{PC}$ conservation reduces allowed 
$ p {\bar p}$  final states 
to several  partial waves. These (denoted by $^{2I+1~2S+1}L_{J}$)  
differ by  isospin $I$, spin $S$, angular 
momentum $L$ and  total spin $J$. 
A different threshold behaviors of  $ p {\bar p}$ 
scattering amplitudes is 
expected in different states.  
Three  partial waves are allowed  in reaction (\ref{1}). 
Two  states   $  ^3P_0 $ and/or $ ^1S_0 $  are preferred by the  angular 
distribution of  photons, but a transition to  $  ^3P_1 $ wave 
is also possible ~\cite{bai03,loi05}. 
Two waves $^{31}P_1$ and $^{33}S_1 $ are possible in the 
$ J/\psi  \rightarrow \pi^0 p\bar{p}$ channel.  

One expects the $ p\bar{p}$ interactions to dominate 
the final state which  becomes an effective two body channel. 
The transition amplitude from an initial  channel $i$ 
to a two-body channel $f$ may be 
presented as 
\begin{equation}
\label{2} 
T_{if}= \frac{A_{if}  }{ 1   +   i  q  A_{ff} }
\end{equation}
where $A_{if}  $ is a transition length,  $A_{ff}  $ is the scattering 
length in the channel 
$f$,  and $ q $ is the momentum in this channel.
The scattering amplitude in  channel $f$ is  given by  
\begin{equation}
\label{3} 
T_{ff}= \frac{A_{ff}  }{ 1   +   i  q A_{ff}}. 
\end{equation}
In the process of interest the formation amplitude $A_{if} $ is  unknown, 
but  $ A_{ff}$ is calculable in $ N \bar {N}$ interaction  models  
constrained  by  other experiments. For slow $ p {\bar p} $ pairs 
one expects $A_{if} \sim q^{L}$ and
$A_{ff} \sim q^{2L}$.
Thus the quantity 
$C_{if} \sim  A_{if}  q^{L}/  A_{ff}$   may be weakly energy dependent,  
which is the essence  of Watson-Migdal approximation 
$ T_{if}\approx~const~\times~q^{-L}~T_{ff}$. 
It  is frequently  true in a  small energy range   
where  the denominator in Eq.~(\ref{2}) provides all the energy
dependence. 
In the  $ p {\bar p}$ states  such an approximation is correct 
for $q$  up to about  0.5~fm$^{-1}$. It fails at  higher momenta 
since   $ A_{ff} $ is  energy dependent  as a result of 
$\pi$ exchange forces.
This has been  pointed out in  Ref.~\cite{sib04} on the basis of an
one-boson exchange version of Bonn potential. A similar behavior 
is seen with the Paris  model~\cite{loi05} although these two 
potentials  differ  
strongly in the two-pion sector. 
On the other hand,  $ A_{if} $ stems  from a  short range $c\bar{c}$ 
annihilation process. The annihilation range  is of  the order 
of $1/m_{c}$~\cite{kro98} and only a  weak energy dependence is  expected 
in $ A_{if} $.  We assume  
$ A_{if} = A^{0}_{if} q^L /(1 + (r_o q)^2 )^{L+1} $
with a range parameter $r_o$ well below 1~fm  and  
a  constant $A^{0}_{if}$.

\textit{The results}. 
The phenomenological  $ A_{ff} $  are fairly well determined 
by the scattering data.  
Here, these are calculated in terms of the updated Paris 
$ N\bar{N}$  potential model~\cite{par05}.
The  model itself is fitted to  
3400  $ \bar{p}p,  \bar{n}p$ scattering  data used in the earlier
version~\cite{par99} and it involves the data  from 
the  $\bar{n}p$ scattering   Ref.~\cite{iaz00} and  $ \bar{p}p$ atoms. 
Figures 1 and 2  present the results. Both decays   
find a natural explanation in this  fairly traditional model 
of $ p \bar{p} $ interactions based on $G$-parity  transformation, 
dispersion theoretical  treatment of two-pion exchange and  
semi-phenomenological  absorptive and short 
range potentials. Quasi-bound states 
close to the threshold are predicted 
 in $ p \bar{p} (^{13}P_1)$,  $ p \bar{p} (^{11}S_0)$ waves 
and a resonance in the $ p \bar{p} (^{13}P_0) $  wave. The first two 
indicate a  strong dependence on the model parameters. 
The third one, the  resonant state, is well established~\cite{got99,car01}.
In order to  see better  the nature of this predictions 
one should look directly  under the $p \bar{p} $ threshold.
An analysis of  the low-energy $ \bar{p}d $ 
scattering or $ \bar{p}d $ atoms allows that, at least in principle. 
Next section discusses  chances to achieve  that.


\section{Subthreshold  amplitudes extracted from ${\bar p}$ atoms }

\begin{table}
\caption{Level shifts in antiprotonic deuterium and He, 
[keV] for $1S$, [eV] for $P$ states. Third column  gives the extracted 
scattering lengths and volumes.}
\begin{tabular}{lll}
          level         & $ \Delta E-i\Gamma/2 $     &   $ A(L)[fm^{2L+1}]$          \\
D, $ 1S$                  & 1.05(25)-i0.55(37) \cite{aug99}  & 0.71(16)- i0.40(27)         \\
D, $ S $  &              scattering  \cite{obe}          &         - i0.62(7) \\
\hline 
D, $2P  $                    & 243(26) -i245(15)    \cite{aug99}     & 3.15(33) - i3.17(19) \\
$^3$He, $2P$                 & 17(4) -  i25(9)    \cite{sch91}     & 4.3(1.0) - i6.3(2.2) \\
$^4$He, $2P$                 & 18(2) -  i45(3)    \cite{sch91}     & 3.5(0.4) - i8.8(1.0) \\

\end{tabular}
\label{table1}
\end{table}

\begin{figure}
\includegraphics[height=.3\textheight]{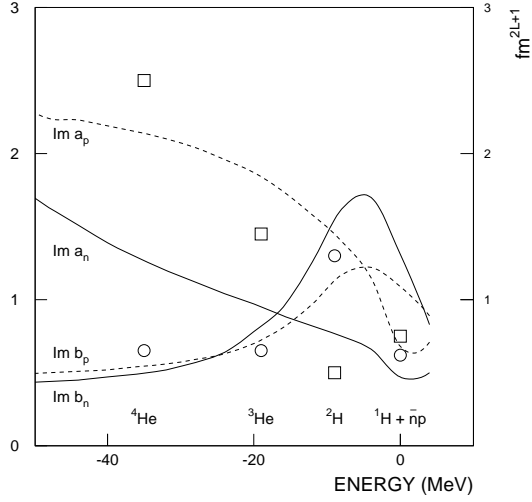}
\caption{The absorptive parts of 
subthreshold amplitudes calculated with Paris  model~\cite{par05}:  
 dotted lines - $\bar{p}p$, continuous lines - $\bar{p}n$ . 
The lengths are denoted as   $ a_p , a_p$  the   volumes 
$ b_p,b_p$. 
 The $b_p/2 + b_n/2$ should be 
compared   to the circles  which give  the average scattering  volumes 
extracted from $d$, $^3$He and $^4$He.  
In the same way  the extracted scattering lengths
given by the squares  are to  be compared to $(a_p + a_n)/2$.
The data support the possible existence of 
$ p \bar{p} (^{33}P_1) $
quasi-bound state found in this model.
In the $S$ wave case, the  statistically insignificant $^{11}S_0$ 
state is possible but not clearly seen.} 
\end{figure}

Experiments which  detect the X-rays emitted 
from hadronic atoms provide electromagnetic levels 
shifted and widened by  nuclear interactions. 
For a given $n$-th state of angular  momentum $L$  
these  complex  level shifts $ \delta E_{nL} - i\Gamma_{nL}/2 $ 
are closely related to the threshold 
scattering parameters $ A_L$,~\cite{lam70},  
\begin{equation}
\label{a1} 
\delta E_{nL} - i\Gamma_{nL}/2= 
\epsilon_{nL}^{o} \frac{4}{n} \Pi_{i=1}^{L}
 ( \frac{1}{i^2}-\frac{1}{n^2} )\frac{A_{L}}{B^{2L+1}}
(1- \lambda A_{L}/B^{2L+1}).   
\end{equation} 
Eq.~(\ref{a1})  is  an expansion in  $ A_L/ B^{2L+1}$
which is  small in all accessible states ($ B$ is the Bohr radius). 
The contemporary precision 
of the experiment requires $1S$ state correction  ($\lambda =3.154$)  
to be included ( few $\%$), otherwise it is negligible. 
The numbers  given in table 1  follow from Eq.~(\ref{a1}) and 
correspond to an average over 
the unresolved fine structure.

In  ${\bar p} d$ the   scattering lengths and volumes may be calculated  
quite reliably by a summation of the multiple scattering series.  
For a full  explanation of the method  we refer 
to~\cite{etad}, it compares 
successfully with exact calculations. Here it is extended to the 
$L=2,3$ states  in  ${\bar p}$ He. 
At zero  ${\bar p}$ energies there are four basic ${\bar p} N$  
amplitudes of interest 
\begin{equation}
\label{a2} 
f_{{\bar p} N}(E)  = a_N(E)  +  3 b_ N(E) {\bf p}\cdot{\bf p}'
\end{equation}
where $ N$ stands for the proton or neutron, 
 ${\bf p} $ and ${\bf p'} $ are the initial and final ${\bar p}N $ CM momenta.
In each case the lengths $a$ and volumes $b$  are averaged over spin states. 
In deuterium ( and other light nuclei) 
 these amplitudes appear to, a good approximation,  
via  the energy averaged values     
\begin{equation}
\label{a3} 
\bar{a}_ N =\int a_ N (-E_B -\frac{p^2}{2 m_{rec}} )\mid
\tilde{\phi}_d^L(p)\mid^2 d\vec{p},  
 \end{equation}
which reflect the  nucleon binding $E_B$ and  the recoil 
of the spectator. The volumes are averaged in a similar way.  
For $d $, the extent of the involved energies is determined by 
the Bessel transforms of the   wave function 
\begin{equation}
\label{a4} 
 \tilde{\phi}_d^L(p) = \int \psi_(r)j_L(pr/2) r^2 dr  
\end{equation}
These  energies  cover some unphysical  subthreshold region. 
The relevant distributions given by  Eq.~(\ref{a3}) peak   
around  -12 and -7  MeV for $ L = 0,1 $ states. 
For heavier nuclei   and stronger nucleon bindings    
the energies of interest are shifted  further away  from the threshold.
That gives the chance to study the energy dependence of $ \bar{a}(E)$ 
and $ \bar{b}(E)$. 

The $\bar{p}$-nucleus scattering parameters may be expressed 
in terms of these averages $A_L(\bar{b},\bar{a})$ by summation 
of the multiple scattering series.
The data  consists  of $1S,2P$ widths + shifts 
in Deuterium,  $2P$ width + shifts  and $3D$ widths 
in $^3$He, $^4$He. With four basic energy dependent  parameters 
$a,b$ a unique resolution is not possible. 
A best fit result may be obtained for  Im $a(E)$ 
and Im $b(E)$. In this 
case additional data from  the $\bar{p}$ stopped 
in $d$  and He chambers~\cite{BIZ74,BAL89}  allow to disclose 
the isospin content of the  absorptive  amplitudes.   
The results are summarized in Fig.~3 and  compared with the  
updated Paris model calculation~\cite{par05}. 
A good 
understanding of the data is obtained. 
Two findings are of interest. First there is an enhancement of the $P$-wave 
absorptive amplitude just below the threshold. Within the model it 
corresponds to a quasi-bound  $^{33}P_1$ state. 
Second, there is an increase of the $S$ wave 
absorption down below  the threshold. Both these effects are fairly well 
understood in terms of the model, although the threshold result 
should be improved.

\textit{Nuclear states of antiprotons} are expected to be very 
 broad and thus difficult to detect. High  angular momentum states 
are narrow, have been seen indirectly in atoms~\cite{klo04}, 
but otherwise are difficult to produce. 
An optimal choice seems to be  a search for a $P$  state in  in the 
reaction  
$^4$He $+ \bar{p} \rightarrow  \bar{p} ^3$H$(^{3}P_0)+p $. 
The indicated  $ ^{3}P_0 $ state  is  the lightest nuclear analogue 
of the  $ ^{13}P_0 $  resonance in  $ p \bar{p}$ system.  It is 
likely to be generated  by the long tail of $\pi$ exchange force 
supplemented by the  Coulomb and core 
interactions.  In the suggested process,  
the final proton energy distribution  would consists of 
a broad structure due to  $S$-wave 
$^3$H-$\bar{p}$   interactions. On top of that,  Eq.~(\ref{2}) 
produces  a narrow $P$-wave quasi-free structure  
given by  $ \mid q/( 1+i q^3 A_1 )\mid^2 $. 
The  averaged $^3$H-$\bar{p}$ scattering   volume     
from table 1 leads to a few MeV wide peak.  
In addition, a few MeV below the $^3$H $\bar{p} $ threshold  one 
would expect a several MeV wide peak  corresponding to the 
nuclear $ \bar{p} ^3$H$(^{3}P_0)$ state. 




\begin{theacknowledgments}
We acknowledge helpful discussions with B. El-Bennich, P. Kienle, 
W. Kloet, M. Lacombe and J.M. Levy. The LPNHE is an Unit\'e de Recherche des Universit\'es Paris 6 et Paris 7, associ\'ee au CNRS. This work was performed in the framework of the IN2P3-Polish Laboratories Convention.
\end{theacknowledgments}

\bibliographystyle{aipproc}   
\IfFileExists{\jobname.bbl}{}
 {\typeout{}
  \typeout{******************************************}
  \typeout{** Please run "bibtex \jobname" to optain}
  \typeout{** the bibliography and then re-run LaTeX}
  \typeout{** twice to fix the references!}
  \typeout{******************************************}
  \typeout{}
 }

\end{document}